\documentclass[twocolumn,tighten,times,fleqn]{aastex62}
\usepackage{color}
\usepackage{multirow}
\usepackage{chngcntr}
\usepackage{mathtools}
\usepackage{amsmath}
\usepackage{amsfonts}
\usepackage{amssymb}
\usepackage[utf8]{inputenc}  
\usepackage{graphicx}        
\usepackage{perpage}         
\usepackage{bm}
\usepackage{lineno}
\MakePerPage{footnote}

\graphicspath{{./}{figures/}}

\makeatletter

\newcommand{\Rmnum}[1]{\expandafter\@slowromancap\romannumeral #1@}
\makeatother

\newcommand{\Oiii}{[O\,{\textsc III}]}
\newcommand{\Oii}{[O\,{\textsc II}]}
\newcommand{\Sii}{[S\,{\textsc II}]}
\newcommand{\Nii}{[N\,{\textsc II}]}

\newcommand{\Oabund}{12+$\log$(O/H)}

\shorttitle{Updated indicators of oxygen metallicity for high-$z$ galaxies}
\shortauthors{Liu et al.}

\begin{document}

\title{Updated Metallicity Diagnostics for Precision Oxygen Abundance Measurements in High-redshift Galaxies with JWST}

\correspondingauthor{Yu Rong}
\email{rongyua@ustc.edu.cn}

\author{Shihong Liu}
\affiliation{Department of Astronomy, University of Science and Technology of China, Hefei, Anhui 230026, China}
\affiliation{School of Astronomy and Space Sciences, University of Science and Technology of China, Hefei 230026, Anhui, China}

\author{Yu Rong$^*$}
\affiliation{Department of Astronomy, University of Science and Technology of China, Hefei, Anhui 230026, China}
\affiliation{School of Astronomy and Space Sciences, University of Science and Technology of China, Hefei 230026, Anhui, China}

\author{Tie Li}
\affiliation{Department of Astronomy, University of Science and Technology of China, Hefei, Anhui 230026, China}
\affiliation{School of Astronomy and Space Sciences, University of Science and Technology of China, Hefei 230026, Anhui, China}

\author{Yao Yao}
\affiliation{Department of Astronomy, University of Science and Technology of China, Hefei, Anhui 230026, China}
\affiliation{School of Astronomy and Space Sciences, University of Science and Technology of China, Hefei 230026, Anhui, China}

\author{Cheng Jia}
\affiliation{Department of Astronomy, University of Science and Technology of China, Hefei, Anhui 230026, China}
\affiliation{School of Astronomy and Space Sciences, University of Science and Technology of China, Hefei 230026, Anhui, China}

\author{Enci Wang}
\affiliation{Department of Astronomy, University of Science and Technology of China, Hefei, Anhui 230026, China}
\affiliation{School of Astronomy and Space Sciences, University of Science and Technology of China, Hefei 230026, Anhui, China}

\author{Hongxin Zhang}
\affiliation{Department of Astronomy, University of Science and Technology of China, Hefei, Anhui 230026, China}
\affiliation{School of Astronomy and Space Sciences, University of Science and Technology of China, Hefei 230026, Anhui, China}

\author{Zhicheng He}
\affiliation{Department of Astronomy, University of Science and Technology of China, Hefei, Anhui 230026, China}
\affiliation{School of Astronomy and Space Sciences, University of Science and Technology of China, Hefei 230026, Anhui, China}

\author{Huiyuan Wang}
\affiliation{Department of Astronomy, University of Science and Technology of China, Hefei, Anhui 230026, China}
\affiliation{School of Astronomy and Space Sciences, University of Science and Technology of China, Hefei 230026, Anhui, China}

\author{Xu Kong}
\affiliation{Department of Astronomy, University of Science and Technology of China, Hefei, Anhui 230026, China}
\affiliation{School of Astronomy and Space Sciences, University of Science and Technology of China, Hefei 230026, Anhui, China}



\begin{abstract}

Recent work has demonstrated that widely used strong-line oxygen abundance indicators, such as O3N2, $\rm R23$, and $\widehat{\rm R}$, suffer from large uncertainties when applied to high-redshift galaxies. We show that this loss of precision primarily arises because, at fixed \Oabund, galaxies span a wide dynamic range in ionization parameter and nitrogen enrichment. Here we develop updated indicators that explicitly incorporate both effects via the proxies O32 and N2O2. We define ${\rm R}_{\rm u}\equiv \rm R23+\alpha_1 O32+\alpha_2 N2O2$, $\widehat{\rm R}_{\rm u}\equiv \rm \widehat{R}+\beta_1 O32+\beta_2 N2O2$, and ${\rm O}_{\rm u}\equiv \rm O3N2+\gamma_1 O32+\gamma_2 N2O2$, and calibrate \Oabund~as low-order polynomials in each composite indicator. Applied to a JWST sample with $T_{\rm e}$-method abundances, the updated indicators substantially tighten the correlations with \Oabund, boosting adjusted coefficients of determination from $\mathbb{R}^2\lesssim 0$ (classical indicators) to $\mathbb{R}^2\gtrsim 0.5$ for the full sample and to $\sim 0.7$ at $z>2$. The residuals reveal a redshift evolution in the mapping between \Oabund, strong lines, ionization, and nitrogen enrichment, with a pivotal turning point near the cosmic noon ($z\sim 2$). Our calibrations provide a practical, physically grounded path to precise metallicity measurements in the JWST era and a firmer basis for quantifying early chemical enrichment and feedback.

\end{abstract}


\section{Introduction} \label{sec:1}

The gas-phase oxygen abundance, \Oabund, is a fundamental tracer of galaxy evolution. It encodes the integrated history of star formation, feedback, and gas flows, serving as a direct constraint on chemical enrichment and baryon cycling across cosmic time. In the local universe, tight empirical relations\---notably the mass-metallicity relation (MZR) and its star-formation-rate-dependent extension, the fundamental metallicity relation (FMR)\---have provided key benchmarks for models of galaxy growth \citep{Tremonti04,Mannucci10,Zahid14,Maiolino19}. At high redshift, the evolution of \Oabund~directly probes the efficiency of early metal production, gas accretion, and outflows during the peak epoch of cosmic star formation and into the reionization era \citep{Erb06,Steidel14,Steidel16,Shapley15,Strom17,Strom18,Sanders16,Sanders21,Kashino19}.

Robust abundance measurements at low redshift rest on a well-established toolkit. The ``direct'' ($T_{\rm e}$) method, which derives electron temperatures from auroral-to-nebular line ratios (e.g., \Oiii$\lambda4363/\lambda5007$), delivers model-independent abundances when such weak lines are detected \citep{Izotov06,Andrews13,Curti17}. For large statistical samples, however, strong-line indicators based on bright forbidden and Balmer lines are indispensable. Classical diagnostics such as R23 \citep{Zaritsky94,McGaugh91,Kewley02,Kobulnicky04}, O3N2 \citep[e.g.,][]{Pettini04}, and N2 \citep{Pettini04,Marino13,Curti17,Curti20} have been extensively calibrated against $T_{\rm e}$-based abundances, yielding empirical, theoretical, or hybrid relations whose systematic offsets are well characterized \citep{Kewley08}. More recent multi-line Bayesian approaches \citep{Blanc15,Perez-Montero14} and composite ratios like N2S2H$\alpha$ and N2O2 \citep{Perez-Montero09,Perez-Montero14,Dopita16} aim to break degeneracies by simultaneously constraining ionization, nitrogen enrichment, and metallicity.

Extending these low-$z$ calibrations to the high-redshift universe remains challenging. Spectroscopic surveys at $z\sim2\--3$ have established that early star-forming galaxies occupy distinct loci in classical diagnostic diagrams \citep[e.g., BPT;][]{Baldwin81}, exhibiting systematically higher \Oiii/H$\beta$, elevated O32 (\Oiii/\Oii), and evidence for harder ionizing spectra and higher ionization parameters compared to local HII regions \citep{Steidel14,Steidel16,Shapley15,Sanders16,Sanders21,Strom17,Strom18,Kashino19}. These differences drive evolution in the zero-points and slopes of strong-line–metallicity relations, exacerbating known degeneracies (e.g., the double-valued nature of R23) and inflating typical \Oabund~uncertainties to 0.1\--0.4 dex when single-ratio calibrations are applied \citep{Kewley08,Bian18,Curti20,Sanders21}. Although strategies such as recalibrating on high-$z$ analogs \citep{Bian18,Sanders21}, constructing multi-parameter diagnostics \citep{Dopita16,Sanders20}, or employing Bayesian photoionization modeling \citep{Blanc15,Perez-Montero14} can mitigate some biases, substantial residual scatter persists, limiting the precision of individual abundance estimates and hampering consistent comparisons across samples and epochs \citep{Steidel16,Strom18,Sanders21}.

The James Webb Space Telescope (JWST) has opened a new window into nebular spectroscopy during the first billion years. Sensitive detections of \Oiii, \Oii, Balmer lines, and\---in an increasing number of systems\---the temperature-sensitive auroral line \Oiii$\lambda4363$ at $z>6\--10$ now provide direct $T_{\rm e}$ anchors in the reionization era, revealing extreme excitation conditions and line-ratio sequences that diverge markedly from local templates \citep{Endsley23,Sanders23,Trump23,Curti24,Nakajima23,Katz23}. In parallel, nitrogen-sensitive diagnostics (N2, O3N2) show pronounced evolution in both normalization and scatter, indicating systematic shifts as well as significant object-to-object variations in N/O relative to $z\sim 0$ \citep{Masters16,Strom18,Sanders21}.

A physically compelling explanation for the persistent scatter in high-$z$ strong-line metallicity estimates is that, at fixed \Oabund, galaxies exhibit a broad range in both ionization parameter (proxied by O32) and nitrogen enrichment (traced by N2O2). Projecting this multi-dimensional parameter space onto a single line ratio inevitably inflates intrinsic scatter and introduces calibration-dependent biases that grow more severe as interstellar medium conditions depart from those of the local calibrating samples \citep{Kewley02,Nakajima14,Dopita16,Perez-Montero09,Masters16,Sanders21}.

In this work, we construct updated strong-line indicators that explicitly incorporate ionization and nitrogen enrichment. Building on widely used one-dimensional diagnostics ($\rm R23$, $\widehat{\rm R}$, O3N2), we introduce composite indicators that combine each classical ratio with O32 and N2O2, then calibrate \Oabund~as low-order polynomials of these composite quantities against direct-method abundances from a JWST sample. We demonstrate that these updated relations dramatically tighten the correlation with \Oabund, reducing scatter and systematic uncertainty across the high-ionization, variable-N/O regime characteristic of early galaxies. 
The paper is structured as follows: Section\ref{sec:2} describes the JWST sample and the derivation of direct-method abundances. Section\ref{sec:3} evaluates the performance of classical strong-line indicators at high redshift. Section\ref{sec:4} presents the updated composite indicators and their calibrations, highlighting the gains in precision and the evidence for redshift evolution. Section\ref{sec:5} summarizes the main conclusions and their implications for future studies. Throughout, ``$\log$'' denotes ``$\log_{10}$''.

\section{Data}\label{sec:2}

\subsection{Sample}\label{sec:2.1}

We assemble a high-redshift galaxy sample from version 4 of the DAWN JWST Archive \citep[DJA;][]{Valentino25}, a public repository of reduced JWST spectroscopic data processed with MSAExp \citep{Brammer23,deGraaff+25}. The DJA pipeline models the continuum with B-splines and fits emission features with Gaussian profiles to extract fluxes for strong lines including \Oii, \Oiii, \Nii, \Sii, and the Balmer series.

\begin{figure*}[!]
\centering
\includegraphics[angle=0,width=0.85\textwidth]{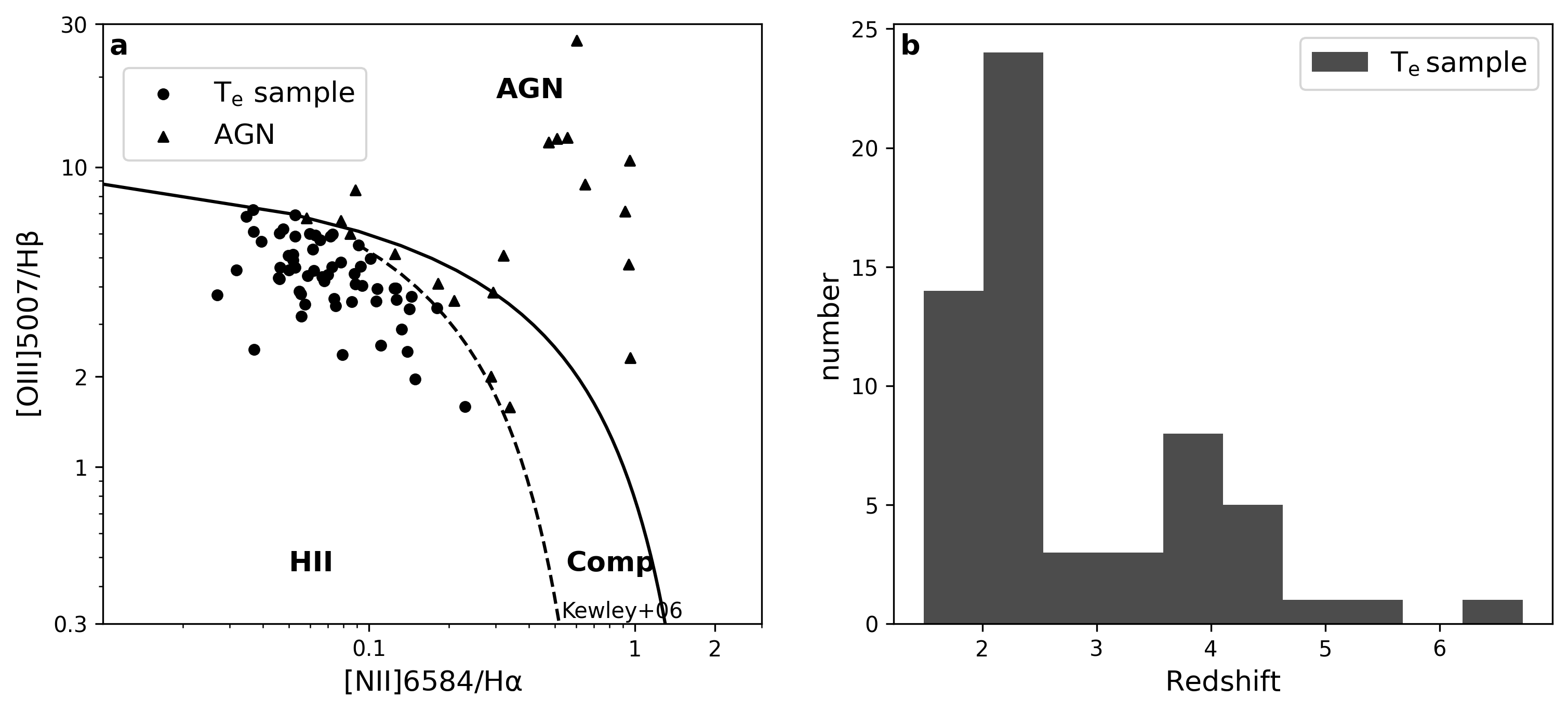}
\caption{a) BPT classification of the DJA sample into star-forming (HII), AGN, and composite systems. b) Redshift distribution of the final high-$z$ star-forming sample (median $\bar{z}\sim 2.24$).}
\label{fig0}
\end{figure*}

JWST spectroscopy in DJA spans three resolving powers, $R\sim 100$, 1000, and 2700. We exclude spectra at $R\sim 100$ because the \Oii$\lambda\lambda3726,3729$ and \Sii$\lambda\lambda6716,6731$ doublets are not cleanly resolved at this resolution. We further require S/N$>3$ in \Oii$\lambda\lambda3726,3729$, \Oiii$\lambda\lambda4959,5007$, \Nii$\lambda6584$, \Oiii$\lambda4363$, and H$\alpha$, H$\beta$. These criteria yield 78 sources. We remove active galactic nuclei, LINERs, and composite systems using the BPT diagram \citep{Kewley01,Kewley06}, resulting in a final sample of 59 star-forming galaxies with a median redshift of  $\bar{z}\sim 2.24$ (Fig.~\ref{fig0}).

Following \cite{Scholte25} and \cite{Cataldi25a}, we define \\
$ \rm R2=\log [\frac{\Oii\lambda\lambda3726,3729}{H\beta}]$ ,\\
$ \rm R3=\log [\frac{\Oiii\lambda5007}{H\beta}]$ ,\\
$ \rm O32=\log [\frac{\Oiii\lambda5007}{\Oii\lambda\lambda3726,3729}]$ ,\\
$ \rm R23=\log [\frac{\Oii\lambda\lambda3726,3729+\Oiii\lambda\lambda4959,5007}{H\beta}]$ ,\\
$ \rm \widehat{\rm R}=\ ${\footnotesize 0.47R2+0.88R3} ,\\
$ \rm O3N2=\log [\frac{\Oiii\lambda5007/H\beta}{\Nii\lambda6584/H\alpha}]$ ,\\
$ \rm N2O2=\log [\frac{\Nii\lambda6584}{\Oii\lambda\lambda3726,3729}]$ ,\\
$ \rm N2=\log [\frac{\Nii\lambda6584}{H\alpha}]$ ,\\
$ \rm N2S2H\alpha = \log [\frac{\Nii\lambda6584}{\Sii\lambda6716,6731}]+0.264\log [\frac{\Nii\lambda6584}{H\alpha}]$.

\begin{figure*}[!]
\centering
\includegraphics[angle=0,width=0.8\textwidth]{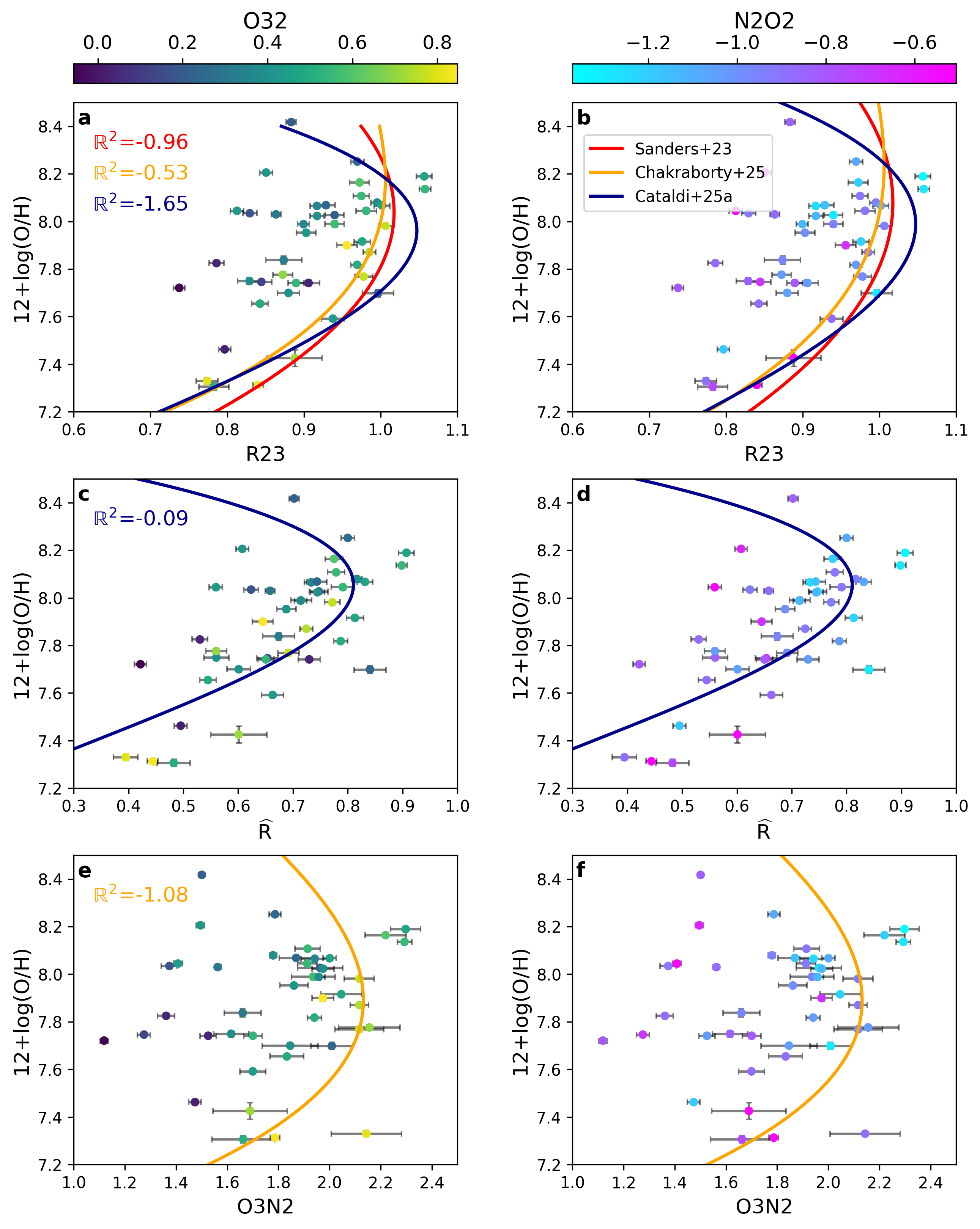}
\caption{Relations between \Oabund\ and classical indicators R23 (top), $\rm \widehat{\rm R}$ (middle), and O3N2 (bottom) for DJA galaxies with direct-method abundances (points). Colored curves denote representative high-$z$ calibrations from the literature. Each panel reports the adjusted coefficient of determination, $\mathbb{R}^2$. Left and right columns are color-coded by O32 and N2O2, respectively.
}
\label{old}
\end{figure*}

\subsection{Direct-method oxygen abundances}\label{sec:2.2}

We derive \Oabund\ using the direct ($T_{\rm e}$) method following \cite{Cataldi25}. Balmer fluxes are corrected for dust attenuation assuming Case B recombination with electron temperature $T_{\rm e}\simeq 10^4$~K and density $n_{\rm e}\simeq 10^2\ \mathrm{cm^{-3}}$, and adopting theoretical decrements H$\alpha$: H$\beta$: H$\gamma$: H$\delta \simeq 2.86: 1.00: 0.468: 0.259$ \citep{Osterbrock06}. We determine $E(B - V)$ via a global $\chi^2$ fit to the available Balmer ratios (S/N$>5$), assuming the \cite{Cardelli89} law; line fluxes are then de-reddened accordingly.

Electron densities $n_{\rm e}$ are estimated from the \Sii$\lambda\lambda$6716, 6731 doublet \citep{Osterbrock06,Kewley19,Sanders16,Tayal10}. For spectra lacking a robust \Sii\ density diagnostic, we adopt $n_{\rm e}\simeq 300\ \rm cm^{-3}$ \citep{Cataldi25}, typical of $z\gtrsim 2$ galaxies \citep[e.g.,][]{Sanders16,Strom17,Stanton25,Topping25a,Shapley15}. This assumption has negligible impact on $T_{\rm e}$ and \Oabund\ within the relevant parameter space \citep{Cataldi25,Osterbrock06,Sanders16,Isobe23}. Uncertainties are propagated via Monte Carlo sampling of the measured line fluxes{\footnote{We note that, the same approach is systematically applied throughout this work to evaluate the uncertainties on all derived parameters.}}.

We assume a two-zone temperature structure. The high-ionization temperature $T_3$ (for O$^{2+}$) is obtained from \Oiii$\lambda4363/\lambda5007$ using PyNeb's getTemDen and the derived $n_{\rm e}$. The low-ionization temperature $T_2$ (for O$^{+}$, S$^+$, and N$^+$) is measured from \Oii$\lambda\lambda7320,7330$ where detected; otherwise we adopt the empirical relation $T_2 = 0.54T_3 + 4790\mathrm{K}$ \citep{Cataldi25a}, valid for $7000$K$<T_3<25000$K. For 43 DJA galaxies, we obtain robust direct-method temperatures and abundances. Ionic abundances O$^{+}$/H$^{+}$ and O$^{2+}$/H$^{+}$ are computed from \Oii/H$\beta$ and \Oiii/H$\beta$ with PyNeb's getIonAbundance, and we take
\begin{equation}
\rm {\frac{O}{H}}={\frac{O^++O^{2+}}{H^+}}.
\end{equation}


\section{Performance of classical indicators}\label{sec:3}

We reassess the \Oabund–indicator relations for R23, $\rm \widehat{R}$, and O3N2 in the DJA subsample with direct abundances (Fig.\ref{old}). 

The indicator R23, a composite of R2 and R3 \citep{Zaritsky94,Kewley02}, has been widely used to estimate \Oabund~in low-redshift galaxies \citep[e.g.,][]{Zaritsky94,McGaugh91,Kewley02,Kobulnicky04}.  
However, as shown in panels a and b of Fig.~\ref{old}, recent studies of high-redshift galaxies reveal that the \Oabund\--$\rm R23$ relationship is weak and exhibits substantial scatter \citep{Cataldi25a,Chakraborty25,Sanders23}. To quantify the goodness of fit, we employ the adjusted coefficient of determination, $\mathbb{R}^2$ \citep{Weisberg80,Greene03}. An $\mathbb{R}^2$ value of, for example, 0.4 indicates that the predictor explains approximately 40\% of the variance in \Oabund, while
$\mathbb{R}^2\lesssim 0$ suggests the model has little to no explanatory power. For our high-$z$ sample, the \Oabund–$\rm R23$ relation yields $\mathbb{R}^2 \lesssim 0$, indicating that R23 is essentially uncorrelated with \Oabund~and would introduce large uncertainties if used for metallicity estimation.

An alternative indicator, $\widehat{\rm R}$, has been proposed to replace R23 for high-$z$ galaxies \citep{Scholte25,Cataldi25a,Laseter24,Chakraborty25}. This metric employs different weightings of R2 and R3 \citep[specifically, $\widehat{\rm R}\simeq 0.47 \rm R_2+0.88 \rm R_3$;][]{Scholte25}  to minimize the $\chi^2$ of the fit, with the relationship often expressed as a cubic polynomial \citep{Scholte25,Cataldi25a}. As shown in panels c and d of Fig.~\ref{old}, while $\mathbb{R}^2$ shows a marginal improvement over R23, its $\mathbb{R}^2$ remains $<0$, indicating that metallicity estimates based on $\widehat{\rm R}$ are still highly uncertain. 
Furthermore, the calibration becomes degenerate for $\widehat{\rm R}>0.8$, falling outside the valid range of the fitted relation \citep{Scholte25}. For galaxies with \Oabund$>7.8$, $\widehat{\rm R}$ exhibits very weak sensitivity, with a correlation coefficient cc$\lesssim 0.3$. These limitations collectively suggest that $\widehat{\rm R}$ is not a robust indicator for high-redshift metallicity.

O3N2 is another commonly used diagnostic \citep{Pettini04,Sanders21,Nakajima22}. However, as indicated in panels e and f of Fig.~\ref{old}, the \Oabund–O3N2 relation for high-$z$ galaxies also shows large scatter and yields $\mathbb{R}^2<0$, implying no meaningful correlation. Other indicators such as N2S2H$\alpha$ and N2 \citep{Scholte25,Dopita16,Steidel14} have similarly been found to perform poorly at high redshift, introducing significant uncertainties in \Oabund~estimates \citep{Scholte25,Cataldi25a}.

Thus, all classical single-ratio indicators introduce substantial uncertainty when applied to high-$z$ galaxies. We posit that the large scatter in the \Oabund–R23, \Oabund–$\widehat{\rm R}$, and \Oabund–O3N2 relations stems primarily from the fact that, at a fixed \Oabund, galaxies exhibit a wide range in ionization parameter (probed by O32) and nitrogen enrichment N/O (probed by N2O2) \citep{Papovich22,Cataldi25,Sanders23,Scholte25,McClymont25}. This is visually confirmed in Fig.~\ref{old}, where galaxies with larger R23, $\widehat{\rm R}$, or O3N2 at a given \Oabund~tend to have higher O32 and lower N2O2. Therefore, to achieve precise metallicity estimates, the ionization parameter and nitrogen enrichment must be explicitly incorporated into the calibration.


\section{Updated indicators of estimating oxygen abundances}\label{sec:4}

To account explicitly for ionization and nitrogen enrichment, we construct composite indicators that augment classical ratios with O32 (ionization proxy) and N2O2 (N/O proxy):
\begin{equation}
\begin{aligned}
& {\rm R}_{\rm u}  \equiv \rm R23+\alpha_1 {\rm O32}+ \alpha_2 \rm N2O2,\\
& \widehat{\rm R}_{\rm u}  \equiv \widehat{\rm R}+\beta_1{\rm O32}+ \beta_2 \rm N2O2,\\
& {\rm O}_{\rm u} \equiv {\rm O3N2}+\gamma_1{\rm O32}+ \gamma_2 \rm N2O2,
\end{aligned}
\label{d_new}
\end{equation}
where $\alpha_1$, $\alpha_2$, $\beta_1$, $\beta_2$, $\gamma_1$, and $\gamma_2$ are free parameters.
We then calibrate \Oabund\ as a low-order polynomial of each composite indicator,
\begin{equation}
12+\log({\rm O/H})=\Sigma^{N}_{i=0} c_ix^i,\ \ x\in \{{\rm R}_{\rm u}, \widehat{\rm R}_{\rm u}, {\rm O}_{\rm u}\},
\label{fit}
\end{equation}
with $N=1$ or 2. 

\begin{figure*}[!]
\centering
\includegraphics[angle=0,width=0.8\textwidth]{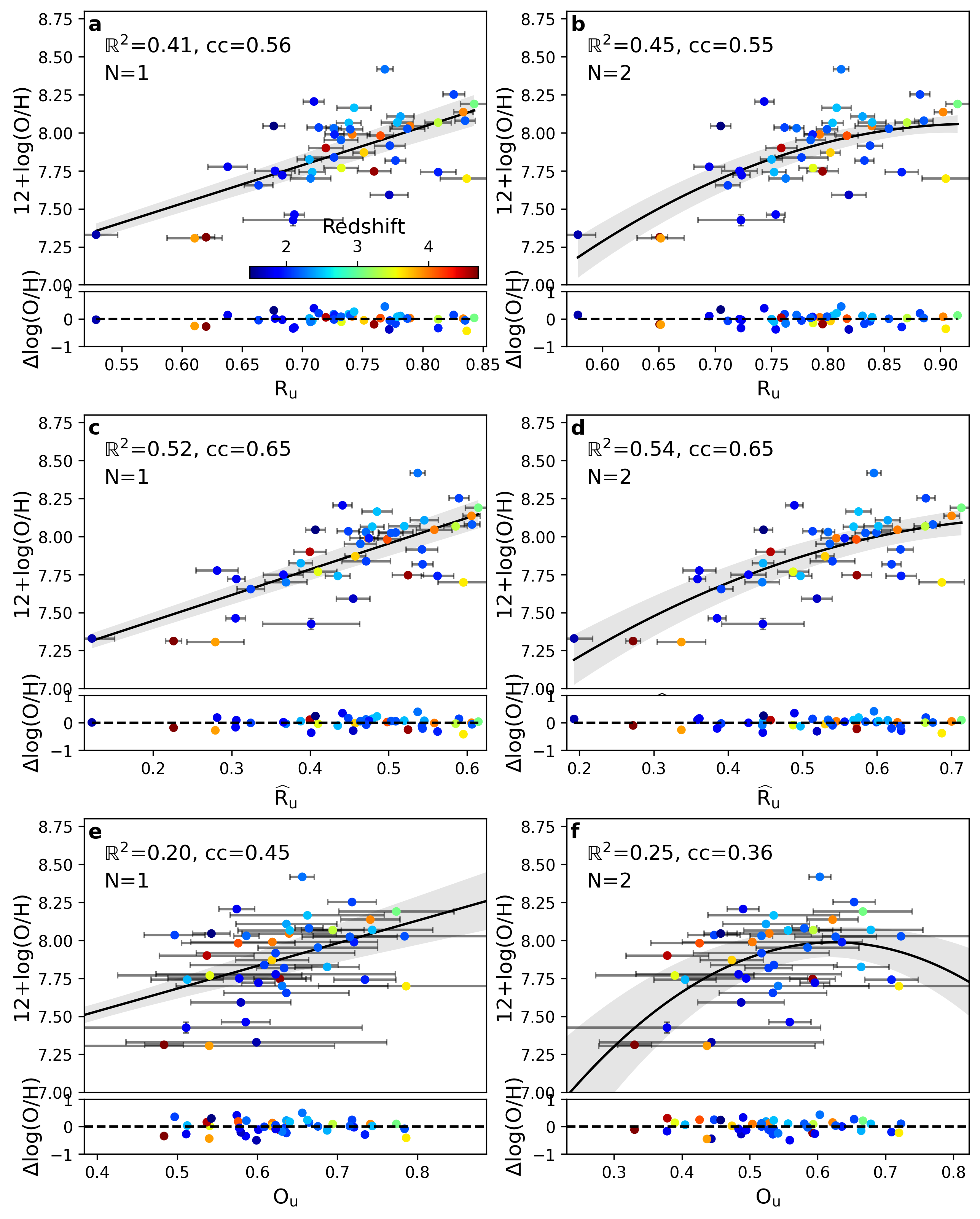}
\caption{Relations between \Oabund\ and the updated indicators $\rm R_{\rm u}$ (top), $\rm \widehat{\rm R}_{\rm u}$ (middle), and $\rm O_{\rm u}$ (bottom). Left and right columns show best-fitting linear ($N=1$) and quadratic ($N=2$) models, respectively. Each panel reports adjusted $\mathbb{R}^2$ and Spearman's cc. Color denotes redshift.}
\label{new}
\end{figure*}

Fitting equations~\eqref{d_new} and \eqref{fit} to the DJA galaxies with direct abundances (Fig.\ref{new}) yields the best-fit parameters of $\alpha_1$, $\alpha_2$, $\beta_1$, $\beta_2$, $\gamma_1$, $\gamma_2$, and $c_i$. We find that, for $x={\rm R}_{\rm u}$:
\begin{equation}
\begin{aligned}
   & \alpha_1=-0.21\pm 0.01, \alpha_2=0.08\pm 0.03,\\
   & c_0=6.02\pm0.11, c_1=2.52\pm 0.21,\\ 
   & ({\rm for}\ \  N=1),\\
    & \alpha_1=-0.21\pm 0.05, \alpha_2=0.03\pm 0.05,\\
    & c_0=1.93\pm 1.15, c_1=13.17\pm 3.04, c_2=-7.07\pm 2.04.\\
    & ({\rm for}\ \  N=2).
\end{aligned}
\end{equation}
For $x=\widehat{\rm R}_{\rm u}$:
\begin{equation}
\begin{aligned}
    & \beta_1=-0.17\pm 0.02, \beta_2=0.16\pm 0.05,\\
    & c_0=7.11\pm 0.04, c_1=1.69\pm 0.12,\\
    & ({\rm for}\ \  N=1),\\
    & \beta_1=-0.15\pm 0.02, \beta_2=0.09\pm 0.06,\\
    & c_0=6.54\pm 0.26, c_1=3.79\pm 0.76, c_2=-2.27\pm 0.77,\\
    & ({\rm for}\ \  N=2)
\end{aligned}
\end{equation}
For $x={\rm O}_{\rm u}$:
\begin{equation}
\begin{aligned}
    & \gamma_1=-1.12\pm 0.61, \gamma_2=0.73\pm 2.17,\\
    & c_0=6.94\pm 0.16, c_1=1.49\pm 0.37.\\
    & ({\rm for}\ \  N=1),\\
    & \gamma_1=-1.29 \pm 2.95 , \gamma_2=0.75 \pm 1.16,\\
    & c_0=5.44 \pm 0.77 , c_1=8.16 \pm 2.02 , c_2=-6.54 \pm1.57.\\
    & ({\rm for}\ \  N=2)
\end{aligned}
\end{equation}


In Fig.~\ref{new}, we show \Oabund~as a function of each indicator with the best-fit parameters. Spearman rank correlations strengthen to cc$\sim$0.55 (\Oabund\---$\rm R_{\rm u}$), cc$\sim$0.65 (\Oabund\---$\rm \widehat{R}_{\rm u}$), and cc$\sim$0.45 (\Oabund\---$\rm O_{\rm u}$), suggesting strong correlation between \Oabund~and the updated indicators. The adjusted $\mathbb{R}^2$ improves to $\mathbb{R}^2\sim 0.41\--0.45$ for ${\rm R}_{\rm u}$, $\mathbb{R}^2\sim 0.52\--0.54$ for $\widehat{\rm R}_{\rm u}$, and $\mathbb{R}^2\sim 0.20\--0.25$ for ${\rm O}_{\rm u}$, demonstrating substantial gains over classical single-ratio indicators.

\begin{figure*}[!]
\centering
\includegraphics[angle=0,width=0.8\textwidth]{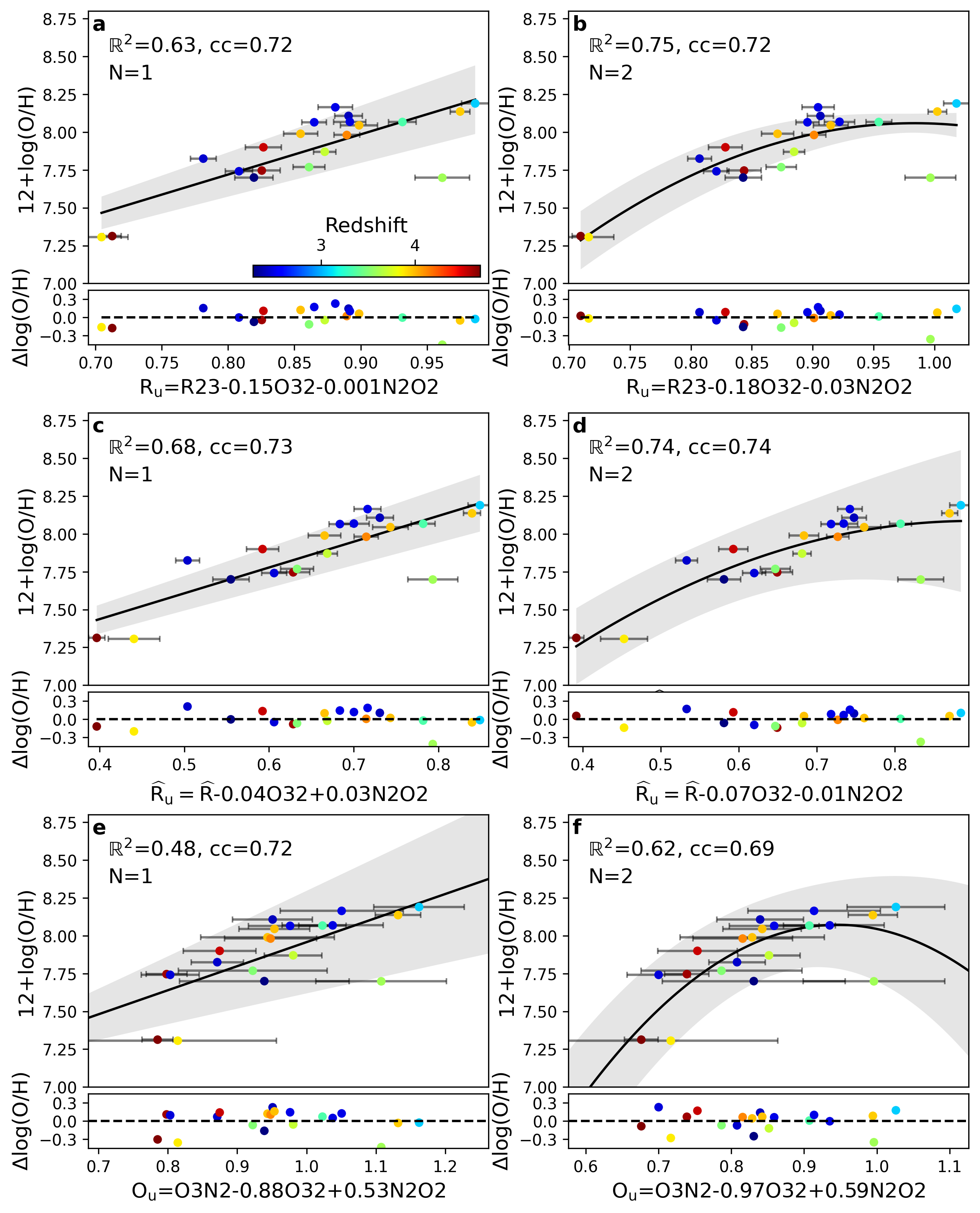}
\caption{Same as Fig.~\ref{new}, but restricted to galaxies with $\bar{z}>2.24$. The relations tighten markedly, with adjusted $\mathbb{R}^2$
reaching $\sim 0.7$ for quadratic models.}
\label{fig4}
\end{figure*}

Splitting the sample at the median redshift $\bar{z}\sim 2.24$ reveals pronounced evolution. As shown in Fig.~\ref{fig4}, for $z>2$ galaxies, the relations tighten further, with adjusted $\mathbb{R}^2$ approaching $\sim 0.7$ for quadratic fits. Best-fit parameters for $z>2$ galaxies are:
\begin{equation}
\begin{aligned}
   & \alpha_1 = -0.15 \pm 0.02, \alpha_2 = -0.001 \pm 0.08,\\
   & c_0 = 5.60 \pm 0.24, c_1 = 2.65 \pm 0.46,\\
   & ({\rm for}\ N = 1),\\
   & \alpha_1 = -0.18 \pm 0.02, \alpha_2 = -0.03 \pm 0.05,\\
   & c_0 = -1.90 \pm 1.58, c_1 = 20.26 \pm 4.28, c_2 = -10.30 \pm 2.77,\\
   & ({\rm for}\ N = 2),
\end{aligned}
\end{equation}
and
\begin{equation}
\begin{aligned}
   & \beta_1 = -0.04 \pm 0.03, \beta_2 = 0.03 \pm 0.09,\\
   & c_0 = 6.75 \pm 0.05, c_1 = 1.70 \pm 0.22,\\
   & ({\rm for}\ N = 1),\\
   & \beta_1 = -0.07 \pm 0.04, \beta_2 = -0.01 \pm 0.13,\\
   & c_0 = 5.48 \pm 0.31, c_1 = 5.77 \pm 0.98, c_2 = -3.20 \pm 1.27,\\
   & ({\rm for}\ N = 2),
\end{aligned}
\end{equation}
and
\begin{equation}
\begin{aligned}
   & \gamma_1 = -0.88 \pm 0.45, \gamma_2 = 0.53 \pm 2.28,\\
   & c_0 = 6.37 \pm 0.28, c_1 = 1.59 \pm 0.60,\\
   & ({\rm for}\ N = 1),\\
   & \gamma_1 = -0.97 \pm 0.07, \gamma_2 = 0.59 \pm 0.13,\\
   & c_0 = -0.25 \pm 1.29, c_1 = 17.60 \pm 3.08, c_2 = -9.31 \pm 2.04,\\
   & ({\rm for}\ N = 2).
\end{aligned}
\end{equation}
At $z<2$, the correlations weaken. The redshift-dependent behavior suggests an evolving mapping among metallicity, ionization, and nitrogen enrichment, with a possible pivot near the cosmic noon.

\section{Summary}\label{sec:5}

Classical strong-line indicators (R23, $\rm \widehat{\rm R}$, and O3N2) perform poorly for high-$z$ galaxies of JWST, yielding $\mathbb{R}^2\leq 0$ and large intrinsic scatter. The dominant driver is multi-parameter mixing: at fixed \Oabund, galaxies span broad ranges in ionization parameter (traced by O32) and nitrogen enrichment (traced by N2O2). By explicitly conditioning on these quantities, our updated composite indicators $\rm R_{\rm u}$, $\rm \widehat{\rm R}_{\rm u}$, and $\rm O_{\rm u}$ markedly tighten the correlations with direct-method abundances, with adjusted $\mathbb{R}^2$ rising to $\sim0.5$ for the full sample and to $\sim0.7$ at $z>2$. These calibrations provide a practical and physically interpretable route to precise metallicity estimates across the high-redshift universe.


\acknowledgments

YR acknowledges supports from the CAS Pioneer Hundred Talents Program (Category B), the NSFC grants 12522302 and 12273037, and the USTC Research Funds of the Double First-Class Initiative. This work is supported by the China Manned Space Program with grant no. CMS-CSST-2025-A06 and CMS-CSST-2025-A08.


%






\end{document}